\documentclass[%
reprint,
longbibliography,
amsmath,amssymb,
aps,
]{revtex4-1}
\usepackage{makecell}
\usepackage{xcolor}
\usepackage{graphicx}
\usepackage{dcolumn}
\usepackage{bm}

\renewcommand{\bf}{\textbf}
\newcommand{\E}{\varepsilon}
\newcommand{\A}{\alpha}
\newcommand{\B}{\beta}

\newcommand{\dbar}[1]{\bar{\bar{#1}}}

\begin{document}
	
	\preprint{APS/123-QED}
	
	\title{Omega-bianisotropic metasurface \\for converting a propagating wave into  a surface wave
		}

	\author{Vladislav Popov}
	\email{uladzislau.papou@centralesupelec.fr}
	\affiliation{%
		SONDRA, CentraleSup\'elec, Universit\'e Paris-Saclay,
		F-91190, Gif-sur-Yvette, France
	}%
	\author{Ana D{\'\i}az-Rubio}
	\author{Viktar Asadchy}
    \author{Svetlana Tcvetkova}
        \affiliation{Department of Electronics and Nanoengineering,
Aalto University, P. O. Box 15500, FI-00076 Aalto, Finland}
	\author{Fabrice Boust}%
    	\affiliation{%
		SONDRA, CentraleSup\'elec, 
		Plateau de Moulon,
		3 rue Joliot-Curie,
		F-91192 Gif-sur-Yvette, France
	}
	    \affiliation{%
		ONERA - The French Aerospace Lab, 91120 Palaiseau, France
	}%
	 \author{Sergei Tretyakov}
	 \affiliation{Department of Electronics and Nanoengineering,
Aalto University, P. O. Box 15500, FI-00076 Aalto,~Finland}
	\author{Shah Nawaz Burokur}%
	    \affiliation{%
		LEME, UPL, Univ Paris Nanterre, F92410 Ville d'Avray, France
	}%

	\begin{abstract}
Although  a rigorous theoretical ground on metasurfaces has been established in the recent years on the basis of the equivalence principle, the majority of metasurfaces for converting a propagating wave into a surface wave are developed in accordance with the so-called generalized Snell's law 
being a simple heuristic rule for performing wave transformations.
Recently, for the first time,  Tcvetkova et al. [Phys. Rev. B 97, 115447 (2018)] have rigorously studied this problem by means of a reflecting anisotropic metasurface, which is, unfortunately, difficult to realize, and no experimental results are available.  
In this paper, we propose an alternative practical design of a metasurface-based converter by separating the incident plane wave and the surface wave in different half-spaces.  
It allows one to preserve the polarization of the incident wave and substitute the anisotropic metasurface by an omega-bianisotropic one.
The problem is approached from two sides: By directly solving  the corresponding boundary problem and by considering the ``time-reversed'' scenario when a surface wave is converted into a nonuniform plane wave.
We  develop a practical three-layer metasurface based on a conventional printed circuit board technology to mimic the omega-bianisotropic response.
The metasurface incorporates metallic walls to avoid coupling between adjacent unit cells and accelerate the design procedure.
The design is validated with full-wave three-dimensional numerical simulations and demonstrates high conversion efficiency.
	\end{abstract}
	
	\maketitle

\section{introduction}

Surface waves  propagate along an interface and exponentially decay away from it being localized on the subwavelength scale. 
Historically, investigation of surface waves started from the discovery of Zenneck waves at radio frequencies and study of optical Wood's anomalies that were explained by the excitation of surface waves~\cite{Maystre_Plasmonics}.
The basic system that supports propagation of surface waves is represented by two semi-spaces filled with a metal and a dielectric~\cite{Maystre_Plasmonics}.
In optical and infrared domains, the effect of strong field localization of surface waves  (or surface plasmon polaritons) is used in many applications only a few of which are listed below.
Specht et al. developed near-field microscopy technique that harnesses surface plasmon polaritons (SPPs) and allows one to significantly overcome the diffraction resolution limit ~\cite{SPP_microscope_Specht1992}.
On-chip SPPs-based high-sensitivity biosensor platforms were implemented and commercialized~\cite{Liedberg1995_SPP_bsed_biosensors,s150510481}.
Surface-enhanced Raman scattering is attributed to excitation of  SPPs~\cite{SPP_review_Zhang2012,SERS2016}.
Application of SPPs in integrated photonic circuits enables further miniaturization in comparison to silicon-based circuits~\cite{Heck2013_Si_PICs} and allows one to approach the problem of size-compatibility with integrated electronics~\cite{Ozbay2006_SPPs_PICs}.

At lower frequencies (THz or microwaves), metals behave like a perfect electric conductor (PEC) what does not allow a surface wave to penetrate in the metallic region but extends it over long distances in a dielectric.
Fortunately, the localization degree can be significantly increased by making use of artificial structures as it was demonstrated in Refs.~\cite{Lee1971,Kildal1990,sievenpiper1999high,Pendry2004_SSPP,Hibbins2005_SSPP,williams2008highly}.
Properties of surface waves excited on a structured interface can be controlled by engineering the interface.
A surface wave propagating along a  periodically structured interface is called as a spoof surface plasmon polariton (SSPPs) and  mimics optical SPPs.
SSPPs allow one to significanty expand the frequency range of SPPs applications.
For instance, SSPPs can be used in integrated microwave photonics~\cite{capmany2007_microwave_photonics,Marpaung2013_microwave_photonics,marpaung2019integrated}.

\begin{figure}[tb]
\includegraphics[width=0.99\linewidth]{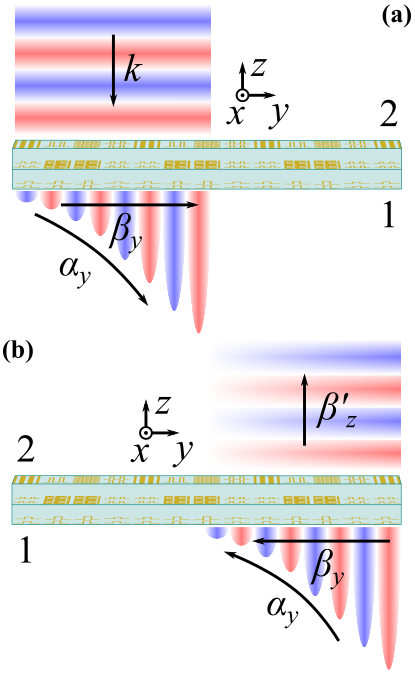}
\caption{\label{fig:1} (a) Schematics of a metasurface converting a normally incident plane wave into a transmitted surface wave with the  propagation constant $\B_y$ and the growth rate $\A_y$. (b) Schematics of a metasurface converting a surface wave into an inhomogeneous plane wave propagating in the normal direction with the propagation constant $\B_z^\prime$.}
\end{figure}

Metasurfaces (or thin two-dimensional equivalent of metamaterials) represent a fruitful tool for manipulation of surface waves~\cite{Capasso2014_review_MSs,Glybovski2016,Chen_2016} and are not restricted to mere support of the propagation of spoof SPPs.
Maci et al. proposed in Ref.~\cite{Maci2011_Metasurfing} a general approach for transforming a wavefront of a surface wave by locally engineering the dispersion relation with spatially modulated metasurfaces.
For instance,  a metasurface-based Luneburg lens for surface waves was demonstrated in Refs.~\cite{Maci2011_Metasurfing,Hibbins2013_Luneburg_lens_for_SEWs}. 
Spatial modulation significantly broadens the range of applications of metasurfaces and  allows one to link propagating waves and surface waves. 
Metasurface-based leaky-wave antennas radiating a surface wave (or more generally, a waveguide mode) into free space were developed in Refs.~\cite{Maci2011_LeakyWave_antenna,Maci2015_Modulated_antenna_space,tcvetkova2018exact,Epstein2019_BLWA}. 
Vice versa, one can take advantage of spatially modulated metasurfaces to convert an incident plane wave into a surface wave [see the schematics in Fig.~\ref{fig:1}(a)], as it was suggested  by Sun et al. in Ref.~\cite{Sun2012}. 
In this case,  an excited surface wave is not an eigenwave and can propagate along a metasurface only under illumination (in contrast to SPPs and SSPPs).
However, one can guide out an excited surface wave on an interface supporting the propagation of the corresponding SSPP~\cite{Sun2012,Sun2016_transmit_HMS_coupler}.
It is worth to note, that metasurface-based converters and leaky-wave antennas are not equivalent, since the plane-wave illumination is normally uniform (other designs also consider Gaussian-beam illumination, see, e.g., Ref.~\cite{Kwon2018_Gaussian_ill}), while a plane wave radiated by a leaky-wave antenna can be essentially inhomogeneous, compare Figs.~\ref{fig:1}(a) and (b). 

Although  a rigorous theoretical ground on metasurfaces has been established in the recent years on the base of the equivalence principle~\cite{Grbic2013,Epstein2014_passive_HMS,Epstein2016_fieldTrans_OBMS}, the majority of metasurfaces for converting a propagating wave into a surface wave are developed in accordance with the so-called generalized Snell's law (see, e.g., Refs.~\cite{Sun2012,Pors2014,Sun2016_transmit_HMS_coupler}).
Initially, the generalized Snell's law was applied to reflect or refract an incident wave at arbitrary angles by engineering the phase of a scattered wave at each point along a metasurface in order to create a linear spacial evolution~\cite{Capasso2011}.
However, in this case the wave impedance of a scattered wave does not equal to the wave impedance of an incident wave. It makes the efficiency of the anomalous reflection (refraction)  to decrease significantly when the angle between the incident and reflected (refracted) wave increases (as well as the impedance mismatch)~\cite{Epstein2014_passive_HMS,Asadchy2016_SpatiallyDispMS,Alu2016_RHMS}.
The outcome is even worse when it comes to the conversion of a propagating wave into a surface wave using the recipe provided by the generalized Snell's law.
The wave impedance of the scattered field is imaginary in this case (a propagating wave has a real wave impedance) and the generalized Snell's law does not  and cannot  ensure a proper energy transfer between the propagating wave and the surface wave (the amplitude of the surface wave must increase along a reactive metasurface  according to the energy conservation, as illustrated by Fig.~\ref{fig:1} (a)).
As a result, losses have to be added to the system in order to arrive at a meaningful solution~\cite{Sun2012}, what makes the generalized Snell's law a tool for designing an absorber rather than a converter (in addition to Ref.~\cite{Sun2012} see also Ref.~\cite{Alu2016_RHMS} where almost perfect absorption is demonstrated by exciting a single near-field mode).

Recently, Tcvetkova et al. have for the first time rigorously studied the problem of conversion of an incident plane wave into a surface wave with a growing amplitude~\cite{Tcvetkova2018} by means of a reflecting anisotropic metasurface (described by tensor surface parameters).
The incident plane wave and the surface wave had orthogonal polarizations in order to avoid interference resulting in the requirement of ``loss-gain''  power flow into the metasurface~\cite{Asadchy2016_SpatiallyDispMS,Alu2016_RHMS}.
Unfortunately, the anisotropic metasurface with the required impedance profile is difficult to realize, and no experimental results are available.  

In this paper, we elaborate on the work done by the authors of Ref.~\cite{Tcvetkova2018} and propose an alternative \emph{practical} design of a metasurface-based converter by separating the incident plane wave and the surface wave of the \emph{same} polarization in different half-spaces. A similar idea was used in Ref.~\cite{Epstein2016_AuxiliryFields} for engineering reflection and transmission of propagating plane waves.  We demonstrate realistic implementation of the converter based on a conventional printed circuit board and confirm its high-efficiency performance via full-wave 3D simulations.  

The rest of the paper is organized as follows. 
In Section II, we derive impedance matrix of a metasurface-based converter.
By means of two-dimensional full-wave numerical simulations, we verify theoretical findings in Section III and propose a topology of a practically realizable metasurface.
Section IV is devoted to description of the design procedure and verification of the design via three-dimensional full-wave simulations.
Eventually, Section V concludes the paper.


\section{Theory}

\subsection{Impedance matrix of an ideal converter}

Consider the conversion of a normally incident plane wave (the magnetic field is along the $x$-axis, see Fig.~\ref{fig:1} (a)) into a transmitted TM-polarized surface wave. Then  the corresponding magnetic and electric fields read as (we assume time-harmonic dependency in the form $e^{i\omega t}$) 
\begin{eqnarray}\label{eq:ansatz_TM}
&&H_{x2}(y,z)=e^{ikz},\quad E_{y2}(y,z)=\eta e^{ikz},\nonumber\\
&&H_{x1}(y,z)=A e^{(\A_z+i \B_z)z}e^{(\A_y-i \B_y)y},\nonumber\\
&&E_{y1}(y,z)=-\frac{i\eta(\A_z+i \B_z)}{k}A e^{(\A_z+i \B_z)z}e^{(\A_y-i \B_y)y}.
\end{eqnarray}
Indices $2$ and $1$ denote the fields above and below the metasurface, respectively, $k$ is the free-space wavenumber, and $\eta$ is the free-space impedance.
All the parameters $\A$ and $\B$ are greater than zero and obey the dispersion relation $(\A_z+i\B_z)^2+(\A_y-i\B_y)^2=-k^2$.
The extinction coefficients $\A_z$ and $\A_y$ result in the surface wave attenuation  away from the metasurface and in its growth along the metasurface (along the $+y$-direction).

We avoid interference between the incident and scattered waves by introducing the latter one  only in the bottom   half-space. Otherwise, the interference would result in complex power flow distribution,   making it difficult to satisfy power conservation conditions locally without gain and lossy structures (also discussed below). 
The chosen configuration when the incident and scattered waves propagate in different half-spaces allows us to   deal with   waves of the same polarization.

\begin{figure*}[tb]
\includegraphics[width=0.99\linewidth]{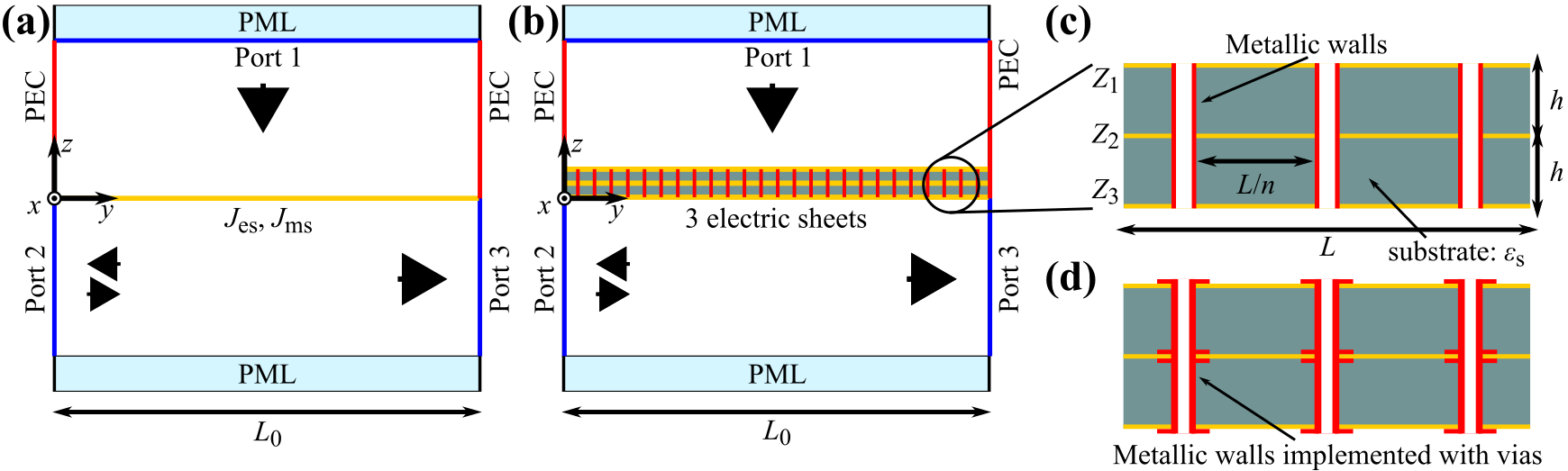}
\caption{\label{fig:2} Schematics of the COMSOL models used for simulating the conversion with (a) omega-bianisotropic combined sheet and (b) asymmetric three-layer structure. 
Port $1$ launches the normally incident plane wave. 
Port $2$ either launches or accepts the surface wave. 
Port $3$ only accepts the excited surface wave. 
(c--d) Zooming of the three-layer metasurface with metallic walls (implemented with vias in (d))
separating individual unit cells, $n$ is the number of unit cells per a super cell, $Z_{i}$ ($i=1,2,3$) is the electric surface impedance of the corresponding sheet.}
\end{figure*}

We characterize the metasurface with a $2\times2$ impedance matrix $\dbar Z (y)$. 
It allows one to understand the most fundamental properties of a system disregarding its concrete physical implementation.  
In terms of an impedance matrix, the boundary conditions determining a metasurface can be written in the following matrix form
\begin{equation}\label{eq:BC}
\left[
\begin{array}{c}
E_{y1}(y,0)\\
E_{y2}(y,0)
\end{array}\right]=
\left[
\begin{array}{cc}
Z_{11}(y)& Z_{12}(y)\\
Z_{21}(y)& Z_{22}(y)
\end{array}\right]
\left[
\begin{array}{c}
-H_{x1}(y,0)\\
H_{x2}(y,0)
\end{array}\right].
\end{equation}
The set of equations~\eqref{eq:BC} serves to find the impedance matrix necessary to perform the transformation given by Eq.~\eqref{eq:ansatz_TM}.
Unfortunately, the desired field distribution Eq.~\eqref{eq:ansatz_TM} does not satisfy these impedance conditions for any reactive  metasurface ($\dbar Z=-\dbar Z^\dagger$, the symbol $\dagger$ stands for the Hermitian conjugate).
The physical reason for this conclusion is that the ansatz fields do not satisfy the energy conservation principle for any choice of the surface-wave parameters~\cite{Tcvetkova2018}.
Although negative, it is an important result: The condition of locally  passive metasurface is a crucial obstacle that does not allow one to perform an ideal conversion of a propagating plane wave into a growing surface wave.
Thus, we omit this  requirement and proceed with a more general impedance matrix $\dbar Z=i \dbar X$, where $\dbar X$ is a real-valued matrix. 
Substituting this ansatz in Eq.~\eqref{eq:BC}, one arrives at the following expression for $\dbar Z$
\begin{eqnarray}\label{eq:Z_PS}
\dbar Z(y)=-i\eta
\left[
\begin{array}{cc}
-\frac{\A_z}{k}+\frac{\B_z}{k}\cot[\B_yy]&
\frac{\B_z}{k}\frac{A\csc[\B_yy]}{\exp[-\A_yy]}\\
\frac{\csc[\B_yy]}{A\exp[\A_yy]}&
\cot[\B_yy]
\end{array}\right].
\end{eqnarray}
Since $X_{12}\neq X_{21}$, the impedance matrix~\eqref{eq:Z_PS} corresponds to a nonreciprocal and locally active or lossy metasurface.
Equation~\eqref{eq:BC} has other then $\dbar Z=i \dbar X$ forms of solutions, as it was shown in~\cite{Tcvetkova2018} for an anisotropic metasurface. 
However, for any exact solution, one arrives at the same conclusion: The impedance matrix corresponds to either reciprocal or nonreciprocal  but always locally active or lossy metasurface.
Noteworthy, active and lossy responses do not necessarily mean that the metasurface must locally radiate or absorb electromagnetic waves. We speculate that a metasurface  possessing strong spatial dispersion can be designed, as it was done in~\cite{Tretyakov2017_perfectAR,8358753} for controlling reflection of propagating waves. 
Unfortunately, the  design procedure of such metasurfaces is still  based on the local periodic approximation~\cite{Pozar1997_LPA,Epstein2016_HMS_review} what does not allow one to set the near-field found a priori~\cite{8358753}.

\subsection{Small growth  approximation}

Conventional leaky-wave antennas  perform  the conversion of a waveguide mode (e.g., a surface wave) into a propagating wave~\cite{balanis2016antenna}.
It makes one think of the reciprocal, ``time-reversed'' process of converting a propagating wave into a surface wave.
We use the quotes to stress that a wave radiated by a leaky-wave antenna is necessarily \emph{inhomogeneous}, while we are particularly interested in converting a \emph{homogeneous}  plane wave into a surface wave. Therefore, these two problems are not equivalent. 
Nevertheless, in practice there are only finite-size antennas and the inhomogeneity can be made arbitrary small (which, however, reduces the radiation efficiency). Let us find the impedance matrix of a metasurface-based leaky-wave antenna converting a TM-polarized surface wave
\begin{eqnarray}\label{eq:ansatz_SW_TM}
H_{x1}(y,z)=A e^{(\A_z-i \B_z)z}e^{(\A_y+i \B_y)y},
\end{eqnarray}
into an inhomogeneous propagating  plane wave with the magnetic field along the $x$-direction 
\begin{eqnarray}\label{eq:ansatz_PW_TM}
H_{x2}(y,z)=e^{-i\B_z^\prime z+\A_y y}.
\end{eqnarray}
Here $\B_z^\prime=\sqrt{k^2+\A_y^2}$ is the propagation constant of the radiated wave.
Figure~\ref{fig:1} (b) depicts a schematics of this process.
The impedance matrix $\dbar Z(y)$ is found by solving the boundary problem formulated in Eq.~\eqref{eq:BC} and becomes symmetric when $A=\sqrt{\B_z^\prime/\B_z}$, thus, corresponding to a reactive and reciprocal metasurface
\begin{eqnarray}\label{eq:Z_SP}
\dbar Z(y)=-i\eta
\left[
\begin{array}{cc}
-\frac{\A_z}{k}+\frac{\B_z}{k}\cot[\B_yy]&
\frac{\sqrt{\B_z\B_z^\prime}}{k}\csc[\B_yy]\\
\frac{\sqrt{\B_z\B_z^\prime}}{k}\csc[\B_yy]&
\frac{k}{\B_z^\prime}\cot[\B_yy]
\end{array}\right].
\end{eqnarray}
Noteworthy, in Ref.~\cite{tcvetkova2018exact} Tcvetkova et al. arrived at a similar impedance matrix for an anisotropic metasurface.
The reader is also directed to Ref.~\cite{Epstein2019_BLWA}, where the authors consider an omega-bianisotropic metasurface-based leaky-wave antenna radiating a waveguide mode that propagates between the metasurface and a ground plane.
In strong contrast with Ref.~\cite{Epstein2019_BLWA}, we employ the concept of leaky-wave antennas as a tool to approach the problem of converting a \textit{uniform} plane wave into a surface wave as discussed further.

\begin{figure*}[tb]
\includegraphics[width=0.99\linewidth]{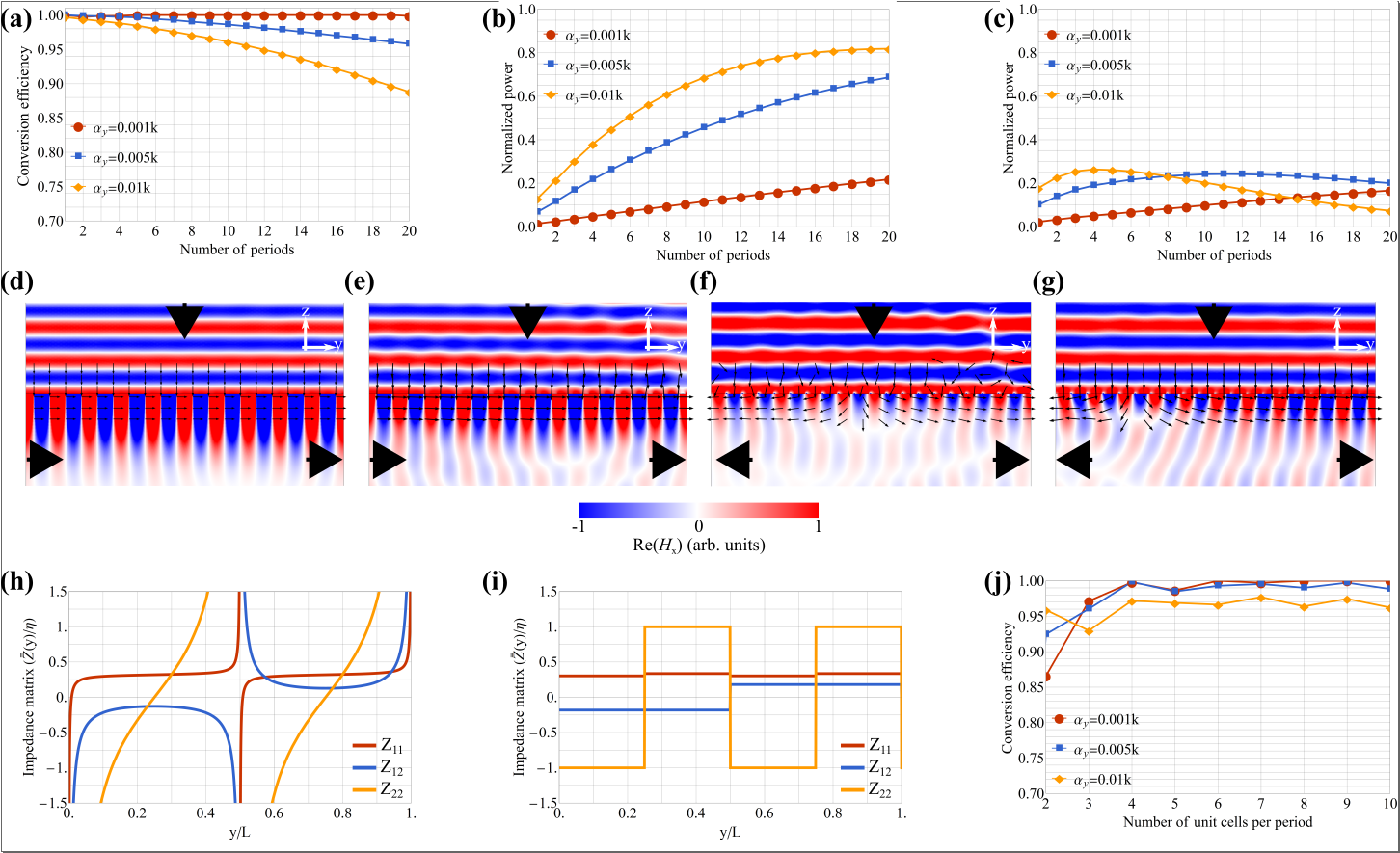}
\caption{\label{fig:3} (a) Conversion efficiency vs. the total length of the metasurface (expressed in terms of the number of periods) for different growth rates $\A_y$ of the surface wave, when the Port 2 is on and excites an input surface wave.
(b--c) Normalized power received by the Ports (b) 3 and (c) 2 vs. the total length of metasurface, when the Port 2 is listening (no input surface wave). 
(d--g) Snapshots of the magnetic field for a metasurface with $10$ periods, the growth rates are (d), (f) $\alpha_y=0.001k$ and (e), (g) $\alpha_y=0.01k$. The Port 2 is on in figures (d--e) and  off in (f--g). The arrows depict directions of the power flow density. 
(h) Continuous and (i) discretized  components (imaginary parts) of the impedance matrix   as functions of the $y$-coordinate. 
(j) Conversion efficiency in case of a discretized impedance matrix vs. the number of unit cells per period (total length of a metasurface is $10L$) for different growth rates $\A_y$ of the surface wave, when the Port 2 is on.
In all figures metasurface is represented by an omega-bianisotropic combined sheet and propagation constant of the surface wave equals $\B_y=1.05k$.
}
\end{figure*}

The reciprocity of the impedance matrix~\eqref{eq:Z_SP} allows one to harness the corresponding metasurface for converting the \textit{inhomogeneous} plane wave at normal incidence~\eqref{eq:ansatz_PW_TM} into the surface wave~\eqref{eq:ansatz_SW_TM}. 
Since we are particularly interested in converting a homogeneous plane wave (this is the case in most practical situations when the source of waves is in the far zone of the metasurface), the total growth of the surface wave amplitude along the length of the metasurface has to remain small.
Mathematically, the small growth condition can be expressed as $\A_yL_0\ll1$, where $L_0$ is the total size of the metasurface in the $y$-direction.
Under the condition $\A_yL_0\ll1$ the impedance matrix~\eqref{eq:Z_SP} (as well as the one given by Eq.~\eqref{eq:Z_PS} when  $A=\sqrt{k/\B_z}$) converges to the following matrix
\begin{eqnarray}\label{eq:Z0}
\dbar Z(y)=-i\eta\left[
\begin{array}{cc}
-\frac{\A_z}{k}+\frac{\B_z}{k}\cot[\B_yy]&
\sqrt{\frac{\B_z}{k}}\csc[\B_yy]\\
\sqrt{\frac{\B_z}{k}}\csc[\B_yy]&
\cot[\B_yy]
\end{array}\right].
\end{eqnarray}
Reactive and symmetric impedance matrix~\eqref{eq:Z0} represents an approximate solution of the boundary problem~\eqref{eq:BC}  and cannot realize \textit{exactly} the transformation represented by Eq.~\eqref{eq:ansatz_TM} even in case of small (but finite) values of the parameter $\alpha_yL_0$. 
Additional waves (not present in Eq.~\eqref{eq:ansatz_TM}) will be excited and play the role of auxiliary waves in the conservation of local normal power flow~\cite{Epstein2016_AuxiliryFields,Tretyakov2017_perfectAR,8358753}. 
Furthermore, in order to satisfy the small growth condition  for a metasurface with the impedance matrix~\eqref{eq:Z0},  an input surface wave should be excited. 
Tcvetkova et al. arrived at the same conclusion in Ref.~\cite{Tcvetkova2018}.
Indeed, the time-averaged power flow density associated with the surface wave in Eq.~\eqref{eq:ansatz_TM} has exponential growth along the metasurface that becomes nearly linear under the small growth assumption (being non-zero along the whole metasurface since $|\A_yy|\ll1$) given by
\begin{equation}\label{eq:S_SW}
\bf S_{SW}(y,0)\approx\frac\eta{2}(1+2\A_y y)\left(\frac{\B_y}{\B_z}\bf y_0-\bf z_0\right).
\end{equation}
In order to create the initial power flow (at $y=0$) along the $y$-direction in accordance with Eq.~\eqref{eq:S_SW}, the amplitude  of the input surface wave  should be equal to $\sqrt{k/\B_z}$ (the amplitude of the excited surface wave~\eqref{eq:ansatz_TM}). 
Vice versa, the amplitude of the excited surface wave will be equal to the one of the input surface wave. 
Since there are two excitation sources (incident homogeneous plane wave and input surface wave), one has to correctly adjust the complex amplitude of the  input surface wave: It must be in phase with that of the incident plane wave and its magnitude must be $\sqrt{k/\B_z}$ times larger.
Only under these conditions nearly all the power of the incident plane wave is transferred to the surface wave. 
Practically, the adjusting procedure can be performed by tuning the power and the phase of the input surface wave (for instance) while measuring the power of the output surface wave. The procedure is over as soon as the maximum of the output power is found.

In spite of all the limitations listed above, the impedance matrix~\eqref{eq:Z0} seems to be the only possible periodic, reactive and reciprocal solution for the conversion problem (which is formulated by the Eq.~\eqref{eq:ansatz_TM} and Eqs.~\eqref{eq:ansatz_SW_TM}, \eqref{eq:ansatz_PW_TM}). 
In what follows, we use only the impedance matrix given by Eq.~\eqref{eq:Z0}.

\section{results of 2D simulations}

In this section we present and analyze results of two-dimensional (2D) full-wave numerical simulations on the conversion of an homogeneous incident plane wave into a surface wave.
In the 2D simulations a metasurface was modeled by means of boundary conditions as described in more detail further.

\subsection{Omega-bianisotropic combined sheet}

\begin{figure*}[tb]
\includegraphics[width=0.99\linewidth]{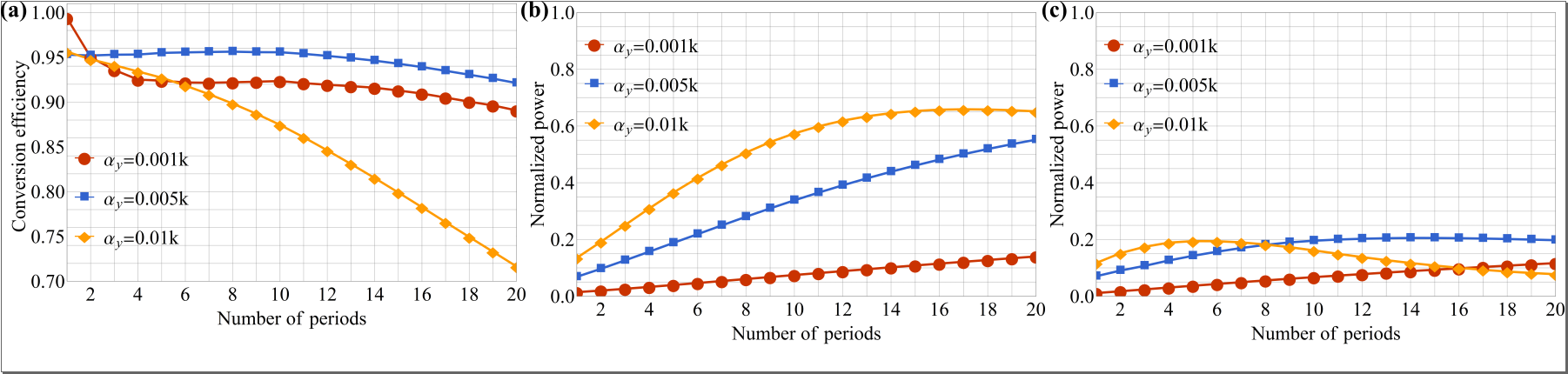}
\caption{\label{fig:6} (a) Conversion efficiency vs. the total length of metasurface (expressed in terms of the number of periods) for different growth rates $\A_y$ of the surface wave, when the Port 2 excites an input surface wave.  (b),(c) Normalized power received by the Ports (b) 3 and (c) 2 vs. the total length of metasurface, when the Port 2 is listening (no input surface wave). 
Metasurface is represented by an asymmetric three-layer structure incorporating metallic walls, the impedance matrix is discretized with four unit cells per period.}
\end{figure*}

A metasurface characterized by a symmetric impedance matrix can be realized as a combined sheet possessing omega-bianisotropic response. 
Then, an incident wave excites electric $J_{\rm es}$ and magnetic $J_{\rm ms}$ surface polarization currents that result in the discontinuity of both tangential electric and magnetic fields at the metasurface. 
In the particular case when the magnetic field is along the $x$-direction, the boundary conditions read as
\begin{eqnarray}\label{eq:BC_surface_currents}
&& H_{x2}(y,0)- H_{x1}(y,0)= J_{\rm es}(y),\nonumber\\
&&E_{y2}(y,0)- E_{y1}(y,0)= J_{\rm ms}(y),\nonumber\\
&&J_{\rm es}=\frac1{Z_{\rm es}}\frac{E_{1y}+E_{y2}}2+K_{\rm me}\frac{H_{1x}+H_{2x}}2,\nonumber\\
&&J_{\rm ms}=Z_{\rm ms}\frac{H_{1y}+H_{y2}}2-K_{\rm me}\frac{E_{1x}+E_{2x}}2.
\end{eqnarray}
Here $Z_{\rm es}$ and $Z_{\rm ms}$ are, respectively, electric and magnetic surface impedances, $K_{\rm me}$ is the magneto-electric coupling coefficient.
When comparing Eq.~\eqref{eq:BC} with Eq.~\eqref{eq:BC_surface_currents}, surface impedances and the coupling coefficient can be expressed in terms of the components of the impedance matrix
\begin{eqnarray}
&&Z_{\rm es}=\frac14\sum_{a,b=1}^2Z_{ab},\quad Z_{\rm ms}=\frac{\det[\dbar Z]}{Z_{\rm es}},\quad K_{\rm me}=\frac{Z_{11}-Z_{22}}{2Z_{\rm es}},\nonumber\\
\end{eqnarray}
where $\det[\dbar Z]=Z_{11}Z_{22}-Z_{12}^2$ is the determinant of $\dbar Z$.

In order to verify theoretical findings and estimate the conversion efficiency, we perform 2D full-wave numerical simulations with \uppercase{COMSOL Multiphysics}. 
The metasurface is represented by  electric and magnetic surface currents set in accordance with Eq.~\eqref{eq:BC_surface_currents}. 
A schematics of the model is illustrated by Fig.~\ref{fig:2}(a). 
Thus, the conversion efficiency is defined as the difference between the output power from Port 3 ($P_3$) and the input power from Port 2 ($P_2$) divided over the power delivered by the incident plane wave from Port 1 ($P_1$): $(P_3-P_2)/P_1$. 

Figure~\ref{fig:3}(a) validates the small growth approximation. 
It is seen that the conversion efficiency approaches $1$ and does not depend on the total length of the metasurface up to $\A_yL_0\sim0.01$.
When increasing the growth rate $\A_y$ (the rest of the parameters are fixed), the conversion efficiency decreases for longer metasurfaces what leads to appearance of spurious scattering in the far-field (compare distribution of the power flow density in  Figs.~\ref{fig:3}(d) and (e)). 

As it was noticed above, the small growth  approximation can be strictly valid only when there is an input surface wave from the Port 2. 
Figures~\ref{fig:3}(b--c) demonstrates the scenario when the Port 2 is listening. 
In bright contrast with the case of Fig.~\ref{fig:3}(a), the part of power of the incident wave coupled to the surface wave increases (but eventually saturates) for larger values of $\A_yL_0$, compare Figs.~\ref{fig:3}(a) and \ref{fig:3}(b).
The difference stems from the normal power flow mismatch at the left end of the metasurface occurring in the case when the Port 2 is switched off.
In the result, surface waves propagating along and opposite to the  $y$-axis  are excited when there is no an input surface wave as demonstrated by Figs.~\ref{fig:3}(b) and (c).
Moreover, it is seen that for small $\A_yL_0$ the power received by the Port 2 is approximately equal to the power received by the Port 3 (and a significant portion of incident power appears in the far-field as spurious scattering).
Snapshots of the magnetic field depicted in Figs.~\ref{fig:3}(f) and (g) show the influence of the spurious scattering on the field profile and power flow distribution in the cases of small ($\A_y=0.001k$) and large ($\A_y=0.01k$) growth rates. 
Specific attention should be paid to the region above the metasurface: Disturbed normally incident power flow indicates the spurious scattering in the far-field.

Although the portion of incident power transfered to the surface wave is considerably higher in case there is an input surface wave, the conversion of a propagating wave into a surface wave usually assumes absence of any input surface wave. At this point one can conclude that metasurfaces do not represent the best approach to the problem but, however, can perform very efficient \textit{enhancement} of an input surface wave (phase an amplitude of the incident plane wave should be accordingly adjusted as discussed in Section II).

Practically, it is important  to study the influence   of the discretization of a continuous impedance matrix on the  performance of a metasurface. The discretized impedance matrix is found from the continuous one as $\dbar Z(y-\textup{mod}(y,L/n)+L/n/2)$, where $n$ is the number of unit cells per period. The components of the impedance matrix as functions of $y$ are plotted in Fig.~\ref{fig:3}(h) for $\beta_y=1.05k$ and $\A_y=0.005k$.
The components  (as functions of $y$) of the corresponding discretized impedance matrix ($n=4$) are shown in Fig.~\ref{fig:3}(i). 
Figure~\ref{fig:3}(j) demonstrates that only in the case of two unit cells per period there is a drop in the conversion efficiency. Making the discretization finer the efficiency quickly grows and reaches the limit of the continuous impedance matrix for as few as $n=4$ unit cells per period. 
This result is very important as allows one to use large unit cell and simplify the design of a sample.

\subsection{Three-layer asymmetric structure}

\begin{figure}[tb]
\includegraphics[width=0.99\linewidth]{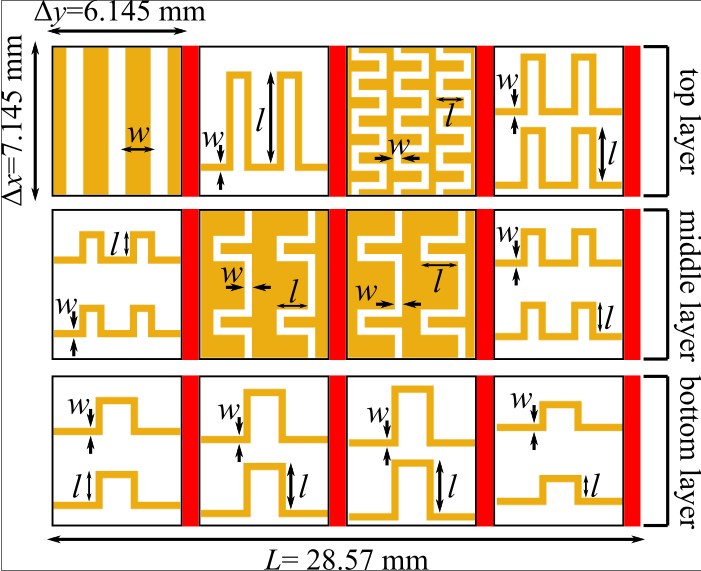}
\caption{\label{fig:7} Topology of the copper (in yellow color) patterns implementing grid impedances in the three-layer design of the metasurface performing the conversion of a normally incident plane wave into the surface wave with $\B_y=1.05k$ and $\A_y=0.005k$ at the frequency $10$ GHz. There are four unit cells per period ($L=2\pi/\B_y\approx 28.57$ mm) separated by metallic walls (illustrated by red rectangles). Thickness of the copper cladding is $35$ $\mu$m. Minimal width of copper traces and gaps is $0.35$ mm.}
\end{figure}

Omega-bianisotropic response can be mimicked with three grid impedances separated by two dielectric substrates~\cite{Epstein2016_fieldTrans_OBMS} as illustrated in Figs.~\ref{fig:2}(b) and (c).
In the COMSOL model grid impedances are introduced via  electric surface currents (in the similar manner with the previous section).
From the transmission line (TL) theory, the impedance matrix~\eqref{eq:Z0} corresponds to the following grid impedances~\cite{Epstein2016_fieldTrans_OBMS} 
\begin{eqnarray}\label{eq:Z_sheets}
&&Z_1=\frac{\eta_{\rm s}\tan(k_{\rm s}h)}{i+\eta_{\rm s}\tan(k_{\rm s}h)\frac{Z_{11}+Z_{12}}{\det[\dbar Z]}},\nonumber\\
&&Z_2=-\frac{(\eta_{\rm s}\tan(k_{\rm s}h))^2\frac{Z_{12}}{\det[\dbar Z]}}{\sec(k_{\rm s}h)^2-2i\eta_s\tan(k_{\rm s}h)\frac{Z_{12}}{\det[\dbar Z]}},\nonumber\\
&&Z_3=\frac{\eta_{\rm s}\tan(k_{\rm s}h)}{i+\eta_{\rm s}\tan(k_{\rm s}h)\frac{Z_{22}+Z_{12}}{\det[\dbar Z]}},
\end{eqnarray}
where $k_{\rm s}=\sqrt{\E_{\rm s}}k$ and $\eta_{\rm s}=\eta/\sqrt{\E_{\rm s}}$, $\E_{\rm s}$ is the relative permittivity of the dielectric substrates (of thickness $h$ each). 
The TL theory assumes that inside the substrates only waves with the $\exp(\mp i k_{\rm s} z)$ spatial dependence propagate.  
This assumption can be strictly valid only for spatially uniform grid impedances.
However, it is not the case of wavefront transforming metasurfaces (and considered metasurface-based converters of propagating waves into surface waves) which require spatial modulation of impedances.
Indeed, closely placed spatially modulated impedance sheets also interact via waves propagating along the substrates which are not  taken into account by Eq.~\eqref{eq:Z_sheets}.
In order to reduce the impact of these waves one has to use very thin and high permittivity  substrates~\cite{Epstein2016_fieldTrans_OBMS}  which refract the waves  closer to the normal direction (and introduce high dielectric losses).
Unfortunately, it still does not allow one to design the grid impedances separately by means of only  Eq.~\eqref{eq:Z_sheets} (due to the coupling between adjacent unit cells).
Instead, Eq.~\eqref{eq:Z_sheets} provides a coarse approximation which is used as a first step of a design procedure aimed at obtaining a given impedance matrix. 

\begin{table}[tb]
\centering
 \caption{\label{tab:1}Physical dimensions of the designed metasurface. Parameter $w$ represents the width of the strip/slot for each inductive/capacitive impedance. Parameter $l$ is the length of the meander in the strips or slots. }
 \resizebox{0.48\textwidth}{!}{%
 \begin{tabular}{|c| c c c c|} 
 \hline
  & Cell 1 & Cell 2 & Cell 3  & Cell 4 \\ [0.5ex] 
 \hline\hline
 Top & \makecell{No meanders  \\  $w=1.23$ mm} & \makecell{2 meanders  \\  $w=0.35$ mm \\ $l=4.48$ mm} & \makecell{4 meanders \\  $w=0.35$ mm  \\  $l=1.35$ mm} & \makecell{2 meanders      \\  $w=0.35$ mm \\  $l=2.65$ mm} \\ 
 \hline
 Mid. & \makecell{2 meanders  \\  $w=0.35$ mm \\  $l=1.23$ mm} &  \makecell{2 meanders  \\  $w=0.35$ mm \\  $l=1.52$ mm} &  \makecell{2 meanders  \\  $w=0.35$ mm \\  $l=1.82$ mm} &  \makecell{2 meanders   \\  $w=0.35$ mm \\  $l=1.52$ mm} \\
\hline
Bott. & \makecell{1 meander  \\  $w=0.35$ mm \\  $l=1.49$ mm} &  \makecell{1 meander  \\  $w=0.35$ mm \\  $l=2.27$ mm} &  \makecell{1 meander  \\  $w=0.35$ mm \\  $l=2.60$ mm} &  \makecell{1 meander  \\  $w=0.35$ mm \\  $l=1.10$ mm} \\
 \hline
 \end{tabular}}
\end{table}

The waves propagating along the substrates can be cut off by means of metallic walls separating each unit cell from the others (in analogy with the idea introduced in acoustics~\cite{Ana_metallic_walls}), see Fig.~\ref{fig:2}(c). 
Practically, metallic walls can be implemented as arrays of vias in a multi-layer printed circuit board.  Such design solution allows one to use substrates of arbitrary large thicknesses $h$ and perform design of a sample considering each grid impedance separately. 
Since a pair of metallic walls represents a parallel plate waveguide inside a unit cell, waves can propagate  with tangential component of wave vector taking the discrete values $\B_m= m \pi/d$ where $d=L/n$ and $m=0,\pm 1,\pm 2,$...
Thus, the finer the discretization,  the less the interaction between the adjacent grid impedances. However, practically it is easier to increase the substrate thickness than to decrease the unit cell size in order to reduce the interaction via higher order spatial harmonics. Figure~\ref{fig:6} demonstrates the dependence of the conversion efficiency on the total length of metasurface and the growth rate $\A_y$ when there is and there is no an input surface wave from the Port 2 [see Fig.~\ref{fig:2}(b)]. 
By comparing Figs.~\ref{fig:3} and \ref{fig:6} one can see that the results for the practical three-layer structure qualitatively repeat those for omega-bianisotropic combined sheet while quantitative  differences are minor and can be explained by the impedance mismatch between the Port 2 and the three-layer metasurface, see Fig.~\ref{fig:2}(b). 

\section{sample design and results of 3D simulations}

\begin{figure}[tb]
\includegraphics[width=0.99\linewidth]{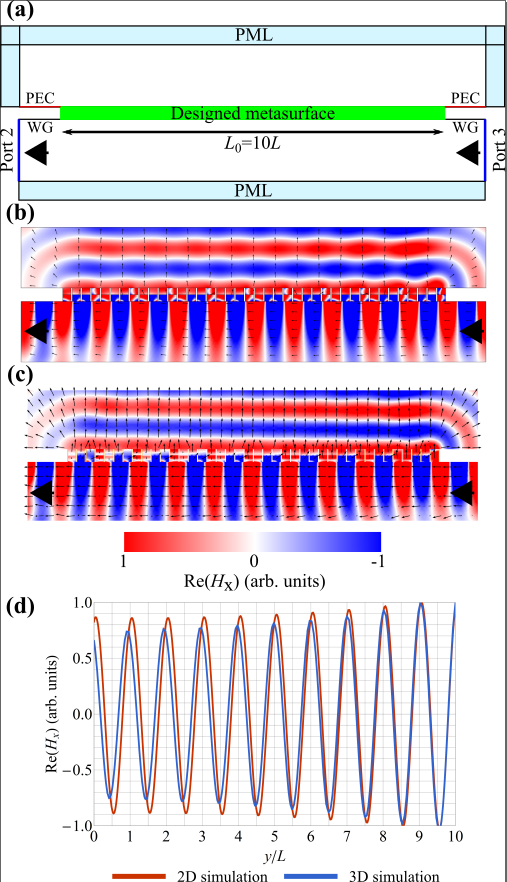}
\caption{\label{fig:8} (a) Schematics of the COMSOL model used for comparing 2D (three-layer metasurface) and 3D (grid impedances are substituted by metallic patterns) simulations, WG section represents surface waveguide implemented as an impedance boundary condition $Z_{WG}=i\eta \A_z/k$.
The Port $2$  accepts the surface wave and the Port $3$ excites an input surface wave. 
(b),(c) Snapshots of the magnetic field for the metasurfaces with $10$ periods in the (b) 2D and (c) 3D simulations, the growth rate is $0.005k$. The arrows depict directions of the power flow density.
(d) Magnetic field along the metasurface (at the distance $\lambda/10$ below the metasurface) extracted from 2D (red curve) and 3D (blue curve) simulations.
}
\end{figure}

The next step towards a real metasurface-based converter is to implement (by means of metallic patterns) three grid impedances found from Eq.~\eqref{eq:Z_sheets}.
The design is performed at the chosen operating frequency of $10$ GHz in accordance with requirements of the conventional printed-circuit-board technology.
On the base of the conducted analysis of 2D simulations, we have chosen the growth rate parameter equal $0.005k$, the propagation constant of the surface wave is $1.05k$.
Eventually, we validate the developed design by comparing the results of 2D and 3D full-wave numerical simulations for a metasurface of total length $L_0=10L$.

The design procedure is based on the commonly used local periodic approximation (see, e.g., Refs.~\cite{Pozar1997_LPA,Epstein2016_HMS_review}). 
Each grid impedance is designed separately.
It is possible due to incorporation of metallic walls and usage of thick dielectric substrates.
Specifically, commercially available F4BM220 substrates with relative permittivity $\E_{\rm s}=2.2(1-i 10^{-3})$ and thickness  $h=5$ mm are used.
The topology of the designed grid impedances is depicted in Fig.~\ref{fig:7}, parameters are specified in Tab.~\ref{tab:1}.

In order to validate the design, we exploit the \emph{reciprocal} scenario  when the metasurface is excited from Port 3 and the Port 2 is listening (Port 1 is absent in this geometry). We compare  2D and 3D simulations.
The schematics of the model is shown in Fig.~\ref{fig:8} (a).
In such a configuration the metasurface transforms the input surface wave from the Port 3 into a propagating wave and becomes a leaky-wave antenna.
Figures~\ref{fig:8} (b) and (c) compare the distribution of the magnetic field obtained in the 2D and 3D simulations, respectively. 
Figure~\ref{fig:8} (d) allows one to see the difference between the magnetic fields at the distance $\lambda/10$ below the metasurface.
Since the metasurface is designed in accordance with the slow growth approximation, not all the power of the surface wave form the Port 3 is launched as a leaky-wave (approximately $50\%$ of power is radiated in the considered example).
Thus, the surface wave entering the Port 2 in Fig.~\ref{fig:8} is the equivalent of the input surface wave in Figs.~\ref{fig:3} and \ref{fig:6}.

\section{discussion and conclusion}

We have theoretically studied  the conversion of a normally incident plane wave into a transmitted surface wave by means of a scalar omega-bianisotropic metasurface.
It allows one to decouple the illumination from the scattered field without changing its polarization and eventually significantly simplifies the design of a sample.
The problem has been approached from two sides: By directly solving  the corresponding boundary problem and by considering the ``time-reversed'' scenario when a surface wave is converted into a nonuniform plane wave.
In agreement with Ref.~\cite{Tcvetkova2018}, we have concluded that  the \textit{perfect} conversion of a \textit{uniform} plane wave into a transmitted surface wave requires the metasurface to exhibit loss-gain response.
On the other hand, a surface wave can be totally radiated into a \textit{nonuniform} plane wave by a reactive reciprocal metasurface.
When imposing the condition of a slowly growing surface wave,  the two approaches lead to the same reactive reciprocal metasurface which can be used for converting a uniform plane wave into a single surface wave with nearly $100\%$ efficiency.
The condition of slow growth requires an input surface wave to create an initial power flow, which is a necessary condition  to have a metasurface with  passive and lossless elements.

The theoretical results have been validated through full-wave 2D simulations by representing a metasurface as a combined sheet with an omega-bianisotropic response.
Next, we have developed a practical three-layer metasurface based on conventional printed circuit board technology  to mimic the omega-bianisotropic response. 
The metasurface incorporates metallic walls to avoid coupling between adjacent unit cells and accelerate the design procedure.
The design has been validated with 2D and 3D simulations and demonstrated high conversion efficiency.
Noteworthy, the three-layer structure is not the only way to achieve the response prescribed by an asymmetric impedance matrix.
Generally, in order to implement omega-bianisotrpic response, one has to consider asymmetric (with respect to the plane $z=0$) unit cells.

As a concluding remark, metasurfaces may not represent the best solution for the matter at hand and other strategies have to be considered.
For instance, a recently emerged concept of metamaterials-inspired diffraction gratings (or metagratings) have demonstrated unprecedented efficiency in manipulating scattered waves with sparse  arrays (contrary to metasurfaces) of polarizable particles~\cite{Alu2017_mtg,Epstein2017_mtg,Popov2019_perfect}.
Due to the sparseness, metagratings inherently possess strong spatial dispersion, what (together with a straightforward design procedure~\cite{Popov2019_LPA}) can be beneficial for solving the conversion problem. 
On the other hand, the near-field of such a  grating is represented by infinite number of modes what makes  it more challenging to selectively excite a given mode.

\bibliography{bib}
 
\end{document}